\documentclass[prl, aps, nofootinbib, twocolumn, amssymb, superscriptaddress, 10pt]{revtex4-2}
\usepackage{amsmath}
\usepackage{amssymb}
\usepackage{amsthm}
\usepackage{amsfonts}
\usepackage{enumerate}
\usepackage{latexsym}
\usepackage{color}
\usepackage{xcolor}
\usepackage{setspace}
\usepackage{blindtext}
\usepackage{dsfont}
\usepackage{mathrsfs}
\usepackage[normalem]{ulem}
\allowdisplaybreaks

\usepackage{lipsum}

\usepackage{tabularx}
\newcolumntype{Y}{>{\centering\arraybackslash}X}

\usepackage{stackengine}
\usepackage{bbold}

\usepackage{bm}
\usepackage{graphicx}

\usepackage{hyperref}
\hypersetup{
pdfnewwindow=true, colorlinks=true,
linkcolor=blue, anchorcolor=blue,
citecolor=blue, filecolor=blue,
menucolor=blue, urlcolor=blue}

\usepackage{pifont}

\newcommand{\beginsupplement}{
        \setcounter{table}{0}
        \renewcommand{\thetable}{S\arabic{table}}
        \setcounter{figure}{0}
        \renewcommand{\thefigure}{S\arabic{figure}}
        \setcounter{equation}{0}
        \renewcommand{\theequation}{S\arabic{equation}}
        \setcounter{section}{0}
        \renewcommand{\thesection}{\Alph{section}}
        \setcounter{subsection}{0}
        \renewcommand{\thesubsection}{\arabic{subsection}}
        \setcounter{subsubsection}{0}
        \renewcommand{\thesubsubsection}{\alph{subsubsection}}
}

\newcommand{\vk}{{\mathbf{k}}}

\newcommand{\vq}{{\mathbf{q}}}

\newcommand{\vlr}{{\mathbf{r}}}

\begin{document}

\title{Topology and compact molecular orbitals
in twisted bilayer \texorpdfstring{WSe$_2$}{WSe2}}

\author{Chenyuan Li}
\thanks{These authors contributed equally to this work.}
\affiliation{Department of Physics \& Astronomy,  Extreme Quantum Materials Alliance, Smalley-Curl Institute,
Rice University, Houston, Texas 77005, USA}
\affiliation{Rice Academy of Fellows, Rice University, Houston, Texas 77005, USA}

\author{Rwik Dutta}
\thanks{These authors contributed equally to this work.}
\affiliation{Department of Physics, University of Texas at Austin, Austin, TX 78712, USA}

\author{Fang Xie}
\thanks{These authors contributed equally to this work.}
\affiliation{Department of Physics \& Astronomy,  Extreme Quantum Materials Alliance, Smalley-Curl Institute, Rice University, Houston, Texas 77005, USA}
\affiliation{Rice Academy of Fellows, Rice University, Houston, Texas 77005, USA}

\author{James R. Chelikowsky}
\affiliation{Department of Physics, University of Texas at Austin, Austin, TX 78712, USA}
\affiliation{McKetta Department of Chemical Engineering, The University of Texas at Austin, Austin, Texas 78712, USA.}
\affiliation{Oden Institute for Computational Engineering and Sciences, The
University of Texas at Austin, Austin, Texas 78712, USA}

\author{Jennifer Cano}
\affiliation{Department of Physics and Astronomy, Stony Brook University, Stony Brook, New York 11794, USA}
\affiliation{Center for Computational Quantum Physics, Flatiron Institute, New York, New York 10010, USA}

\author{Mit H. Naik}
\thanks{\href{mit.naik@austin.utexas.edu}{mit.naik@austin.utexas.edu}}
\affiliation{Department of Physics, University of Texas at Austin, Austin, TX 78712, USA}
\affiliation{Oden Institute for Computational Engineering and Sciences, The
University of Texas at Austin, Austin, Texas 78712, USA}

\author{Qimiao Si}
\thanks{\href{qmsi@rice.edu}{qmsi@rice.edu}}
\affiliation{Department of Physics \& Astronomy,  Extreme Quantum Materials Alliance, Smalley-Curl Institute, Rice University, Houston, Texas 77005, USA}

\date{\today}

\begin{abstract}
     Recent observations of superconductivity in twisted bilayer $\mathrm{WSe_2}$ (t$\mathrm{WSe_2}$) have motivated theoretical proposals for unconventional pairing mechanisms. A central question is whether band topology plays an essential role in the system's correlation physics.
    In this letter, we develop a first-principles-based description of the top moiré valence bands in t$\mathrm{WSe_2}$.
    Using density functional theory (DFT) calculations,
    we identify the bands in the relevant range of twist angles to be topologically non-trivial, with the top valence bands carrying Chern numbers $C=(+1,+1)$ for the $K$ valley.
    In order to treat the strong correlation physics,
    we construct compact molecular orbitals directly from the DFT wave functions through a partial Wannierization procedure and with the guidance of spinful $C_{3z}$ symmetry representations.
    This yields a localized $f$ orbital together with a complementary topological $c$ orbital, allowing us to extract hopping and hybridization amplitudes from  first principles.
    The resulting parameters provide an {\it ab initio} benchmark for the effective Hamiltonian.
    Our work establishes a foundation for understanding superconductivity in moiré TMDs and highlights t$\mathrm{WSe_2}$ as a promising platform for exploring topological superconductivity.
\end{abstract}

\maketitle

{\it \color{blue} Introduction}---
Moiré transition-metal dichalcogenides (TMDs) have emerged as highly tunable platforms for studying the interplay between strong correlations and band topology.
Compared with twisted bilayer graphene, where flat bands occur near special magic angles, twisted TMD homobilayers naturally host narrow bands over a wider range of twist angles due to their large effective masses and strong moiré potentials.
Together with strong spin-orbit coupling and electrostatic tunability, these features allow correlation and topology to be tuned in a highly flexible way.
Experiments have revealed a broad range of correlated phenomena in these materials, such as correlated insulating behavior, Wigner crystals \cite{Wang2020}, and tunable metal-insulator transitions \cite{ghiotto_quantum_2021},   highlighting these systems as solid-state simulators for correlated physics.
More recently, superconductivity has been observed in twisted $\mathrm{WSe_2}$ (t$\mathrm{WSe_2}$) \cite{Xia2024Unconventional, Guo2024Superconductivity, Guo2026, Xia2026Bandwidth}.
Importantly, for the case of e.g. the twist angle
$\theta=3.6^\circ$, experiments have observed clear signatures
of strong correlations, through such features as
a large magnitude of the
high temperature resistivity, reaching the Mott-Ioffe-Regel value,
correlated insulating state in the phase diagram
and indications of quantum criticality.
Several theoretical proposals have been put forward.
Some have been anchored by a strongly correlated normal state,
describing pairing  driven by topology-induced quantum fluctuations
\cite{Xie2025Superconductivity, Xie2025Kondo,Li2025Topological},
while others start from a variety of normal state
\cite{Tuo2025Theory,Qin2024KohnLuttinger,Guerci2024Topological,Christos2024Approximate,Wu2023Pair,Chen2025,Kuhlenkamp_Arxiv_2025}.

A central question in the  basic understanding concerns the extent to which the  topology of the top moiré bands is essential to the correlation physics.
Previous theoretical studies have either neglected band topology or largely assumed the band topology based on a continuum model \cite{Wu2019topological, Pan2020Band, Devakul2021Magic,Zhang2025Twist,Zhang2024universal}.
However, the topology of TMD moiré bands can sensitively depend on a host of factors such
as the twist angle, moiré potentials and lattice relaxation.
For these reasons, even the basic question of whether the relevant moiré bands
are topological has remained open.

In this letter, we develop a first-principles-based effective description of the top moiré valence bands in twisted bilayer $\mathrm{WSe_2}$.
We focus on the range of twist angles that pertain to the superconductivity and compute the {\it ab initio} moiré band structures.
Over the entire twist angle range, we find the top moiré  bands to be topological.
These results are enabled by the DFT methods introduced previously \cite{Kundu2022Moire,Zhang2024Polarization}.
In order to treat the correlation physics, we then construct compact molecular orbitals \cite{chen2023Metallic,hu2023coupled,Chen2024Emergent,souza2026origin}
directly from the DFT wavefunctions through a partial Wannierization procedure:  a maximally localized $f$ orbital captures the compact component, while a complementary $c$ electron carries the remaining topological obstruction.
This provides
a firm foundation for understanding the superconductivity in twisted $\mathrm{WSe_2}$
based on
a correlated topological
description in general, and
an effective Kondo lattice approach in particular.

{\it \color{blue} Band structures and spatial profile}---
\begin{figure}[t]
    \centering
    \includegraphics[width=\linewidth]{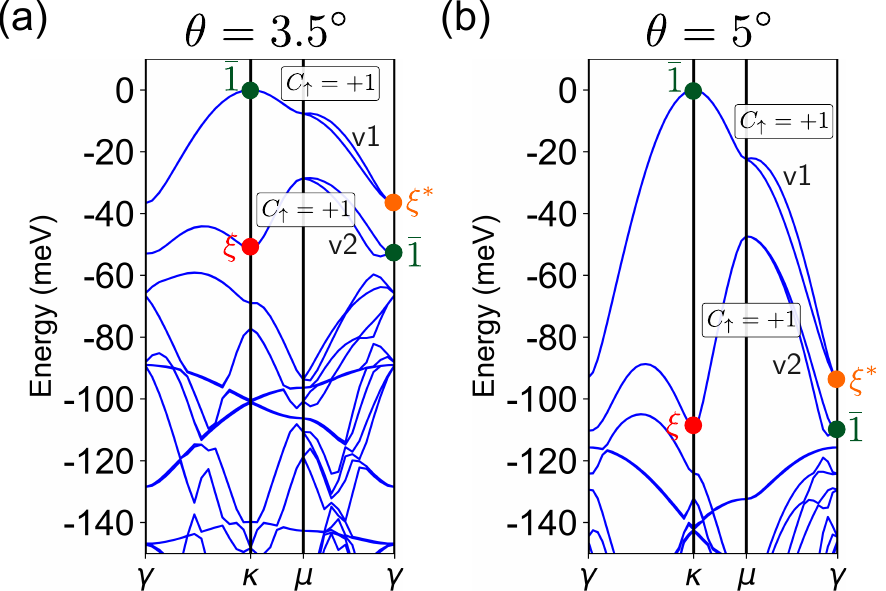}
    \caption{Band structures for $\mathrm{tWSe_2}$ at twist angles (a) $3.5^\circ$ and (b) $5.0^\circ$. The two isolated spin-polarized bands are labeled by their Chern numbers, $C_{\uparrow}=+1$. The labels denote the spinful $C_3$ rotation eigenvalues of the spin up states at the high-symmetry points, with $\xi=e^{i\pi/3}$, $\xi^*=e^{-i\pi/3}$ and $\bar{1}=-1$. The corresponding eigenvalues for spin down states are related by time-reversal symmetry.
    In the top two bands, the $C_{3z}$ eigenvalues for $\kappa$ and $\kappa'$ points are identical for the same spin.
    }
    \label{fig:band}
\end{figure}
We focus on the valence bands of twisted $\mathrm{WSe_2}$ homobilayers and compute their electronic structure by DFT to investigate the topology of moiré bands across a range of twist angles.
As shown in Fig.~\ref{fig:band}, the top valence bands arise from the folded $K$- and $K'$-valley states, which are related by time-reversal symmetry and carry opposite spins.
The folded minibands from the two valleys appear close in energy and exhibit a small splitting, most clearly along the $\gamma$-$\mu$ direction.
This splitting can be attributed to weak intervalley coupling or trigonal warping.

We next characterize the moiré band topology using the spinful $C_{3z}$ rotation eigenvalues of the DFT Bloch wavefunctions.
For a spinful system, the $C_{3z}$ eigenvalues satisfy $(C_{3z})^3=-1$, and we denote them by $\xi=e^{i\pi/3}$,  $\xi^*=e^{-i\pi/3}$ and $\bar{1}=-1$.
The Chern number modulo three can be inferred from the rotation eigenvalues at the $C_{3z}$-invariant momenta of the moiré Brillouin zone (mBZ) through $\exp(i\frac{2\pi}{3}C)=-\xi_\gamma\xi_\kappa\xi_{\kappa'}$, where the $\xi$s are the corresponding spinful $C_{3z}$ eigenvalues.
These eigenvalues are labeled in Fig.~\ref{fig:band}.
We find that the two topmost moiré bands are topologically nontrivial.
In the range of twist angles examined here, $3.5^\circ\leq \theta \leq 5^\circ$, both spin-up $K$-valley bands carry Chern number $C=+1$.
The opposite-spin sector is related by time-reversal symmetry, under which the Berry curvature changes sign and the $C_{3z}$ eigenvalues are complex conjugated; consequently, the spin-down $K'$-valley bands carry the opposite Chern numbers.
Our conclusion is supported by recent calculations at related twist angles \cite{Zhang2024Polarization}.

\begin{figure}[t]
    \centering
    \includegraphics[width=\linewidth]{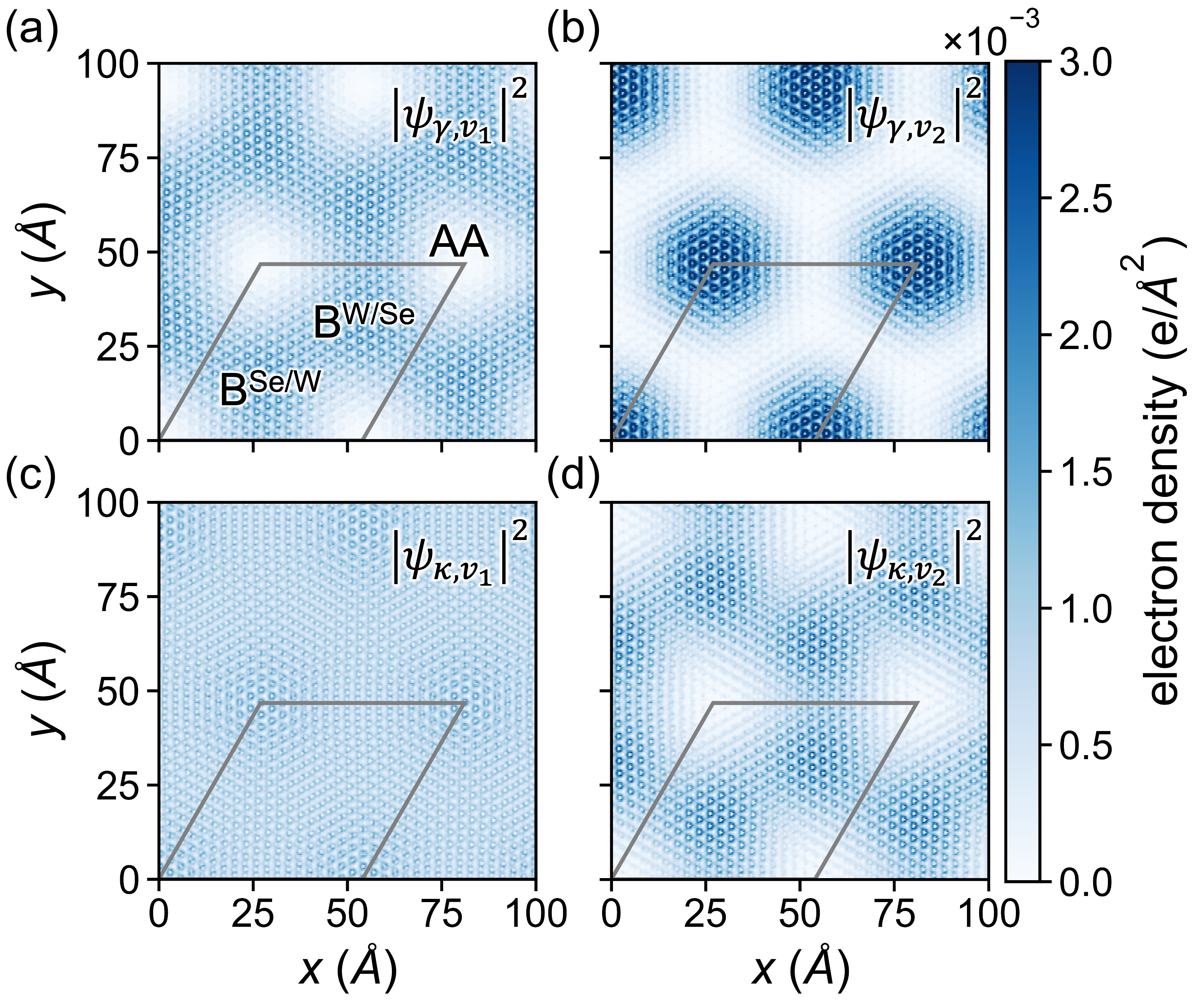}
    \caption{Distribution of Kohn-Sham states $|\psi_{\bm{\gamma}}(\mathbf{r})|^2$ and $|\psi_{\bm{\kappa}}(\mathbf{r})|^2$ averaged along the out-of-plane ($z$) direction for the top two valence bands at twist angle $\theta=3.5^\circ$. Panels (a,b) are evaluated at the moir\'e $\bm{\gamma}$ point, and panels (c,d) at the moir\'e $\bm{\kappa}$ point. Panels (a,c) show the topmost band, while panels (b,d) show the second top band. Dashed parallelgrams mark one moir\'e unit cell. }
    \label{fig:density}
\end{figure}

This result makes it important to take
the band topology into account when studying the correlation physics.
Towards this end, we now examine the real-space distribution of the density corresponding to the top two valence band states at representative $\mathbf{k}$ points in the moiré Brillouin zone.
Fig.~\ref{fig:density}(a) and \ref{fig:density}(b) show the 
spatial profile of the density corresponding to the top two valence bands at the moiré $\gamma$ point, which we denote by V1 and V2 respectively.
The V1 density forms a hexagonal pattern, with suppressed weight near the AA stacking regions and enhanced weight around the AB and BA regions, whereas the V2 state is predominantly localized in the AA regions.
At the moiré $\kappa$ point, shown in Fig.~\ref{fig:density}(c) and \ref{fig:density}(d), the V1 state becomes more extended over the moiré unit cell,  while the V2 density is concentrated around AB and BA regions, forming a hexagonal pattern.
Both spin sectors are included in our analysis.
The spin-up and spin-down wavefunctions exhibit complementary spatial profiles, which are related by time-reversal symmetry.
These localization patterns are also qualitatively in agreement with STM measurements at $3^\circ$ t$\mathrm{WSe_2}$, which observed distinct local density of state patterns associated with different points of the unit cell Brillouin zone \cite{Zhang2020Flat}.

This real space electron distribution provides a direct microscopic connection to the continuum model description and the resulting effective Kondo lattice $t$-$J$ model \cite{Xie2025Kondo,Li2025Topological}
(further details are provided in Sec~\ref{sec:Wannierization_cont}).
The spatial profiles obtained from DFT are consistent with the results of the continuum model shown in Fig.~\ref{fig:cont_density},
where low-energy degrees of freedom are described by the exponentially localized $f$ orbital  hybridized with extended ``clouds'' of the topological power-law-decaying $c$ orbital.
In this basis, the wavefunction weights from the $f$ and $c$ orbitals depend on the momentum.
Near the moiré $\gamma$ point, the V2 wavefunction is dominated by the localized $f$ orbital, while V1 has stronger weight from the more extended $c$ orbital component.
Away from $\gamma$, this orbital character is largely exchanged: V1 becomes predominantly $f$-like, whereas V2 acquires stronger $c$-orbital character.
This momentum-dependent interchange of localized and extended orbital components is consistent with the topological nature of V1 and V2, and supports the Kondo lattice description of twisted $\mathrm{WSe_2}$.

{\it \color{blue} Compact molecular orbital construction from the DFT bands}---
The nontrivial topology of the two topmost moiré bands imposes an important constraint on the construction of localized orbitals.
Because the two bands carry the same nonzero Chern number, they cannot be
Wannierized into exponentially localized orbitals that preserve the $C_{3z}$ symmetry.
This obstruction motivates a DFT-based counterpart of the ``{\it partial Wannierization}'' approach developed in continuum models, which involves a maximally localized Wannier function (MLWF), and a topological power-law orbital (TPLO) carrying the entire topological obstruction of both moiré bands \cite{Xie2025Kondo,Li2025Topological}.
 The spatial extent of the MLWF is comparable to the moiré length scale and much larger than the size of an atomic orbital.
 Therefore, it can be viewed as an effective molecular orbital,  while it remains compact compared with its orthogonal topological counterpart.

The spinful $C_{3z}$ eigenvalues in Fig.\ref{fig:band} provide the symmetry guide for this construction.
For spin up at $\theta=3.5^\circ$, the top-most band has $C_{3z}$ eigenvalues of $\xi^*$, $\bar{1}$, $\bar{1}$ at $\gamma$, $\kappa$ and $\kappa'$ points, respectively, while the second band carries $\bar{1}$, $\xi$, $\xi$  at these three points.
Thus, we seek a Wannier function for a single topologically trivial band with eigenvalues $\bar{1}$ at all three high-symmetry points, which has weight in V1 at most $k$ points including $\kappa$ and $\kappa'$ and in V2 at $\gamma$.
This corresponds to the elementary representation (EBR) induced by the $^{2}\bar{E}$ representation of the $1a$ Wyckoff position ($\bm{r}_{1a} = \bm{0}$) of the space group {\it P}3 \cite{Cano2021band}.
This band inversion near the moiré $\gamma$ point further implies that the two orbitals will hybridize with each other.
The same symmetry-based construction can be applied to spin down or at other twist angles, including $\theta=5^\circ$.

To implement this construction, we perform the disentanglement Wannierization procedure provided by \textsc{Wannier90} \cite{Marzari1997Maximally,Souza2001Maximally,marzari_maximally_2012,Pizzi2020wannier}.
In the following, we focus on the  $K$-valley moiré bands by introducing a small Zeeman field, which separates the spin-up and spin-down sectors without substantially modifying their orbital character or dispersion.
For the one-valley model, the symmetry group is generated by $C_{3z}$.
We therefore take the trial orbital to be centered at the $1a$ Wyckoff position and to transform according to the $C_{3z}$ representation inferred from the symmetry eigenvalues of the DFT bands.
From this symmetry-adapted trial state, the compact $f$ orbital is constructed  by \textsc{Wannier90}, which selects the gauge that minimizes the real space spread of the resulting orbital.
Once the $f$ orbital is fixed, the remaining $c$ orbital can be obtained as the orthogonal complement within the two-band Hilbert space at each $\vk$.
The technical detail of this approach is outlined in Sec.~\ref{sec:Wannierization_DFT} in the SM.

\begin{figure}[t]
    \centering
    \includegraphics[width=\linewidth]{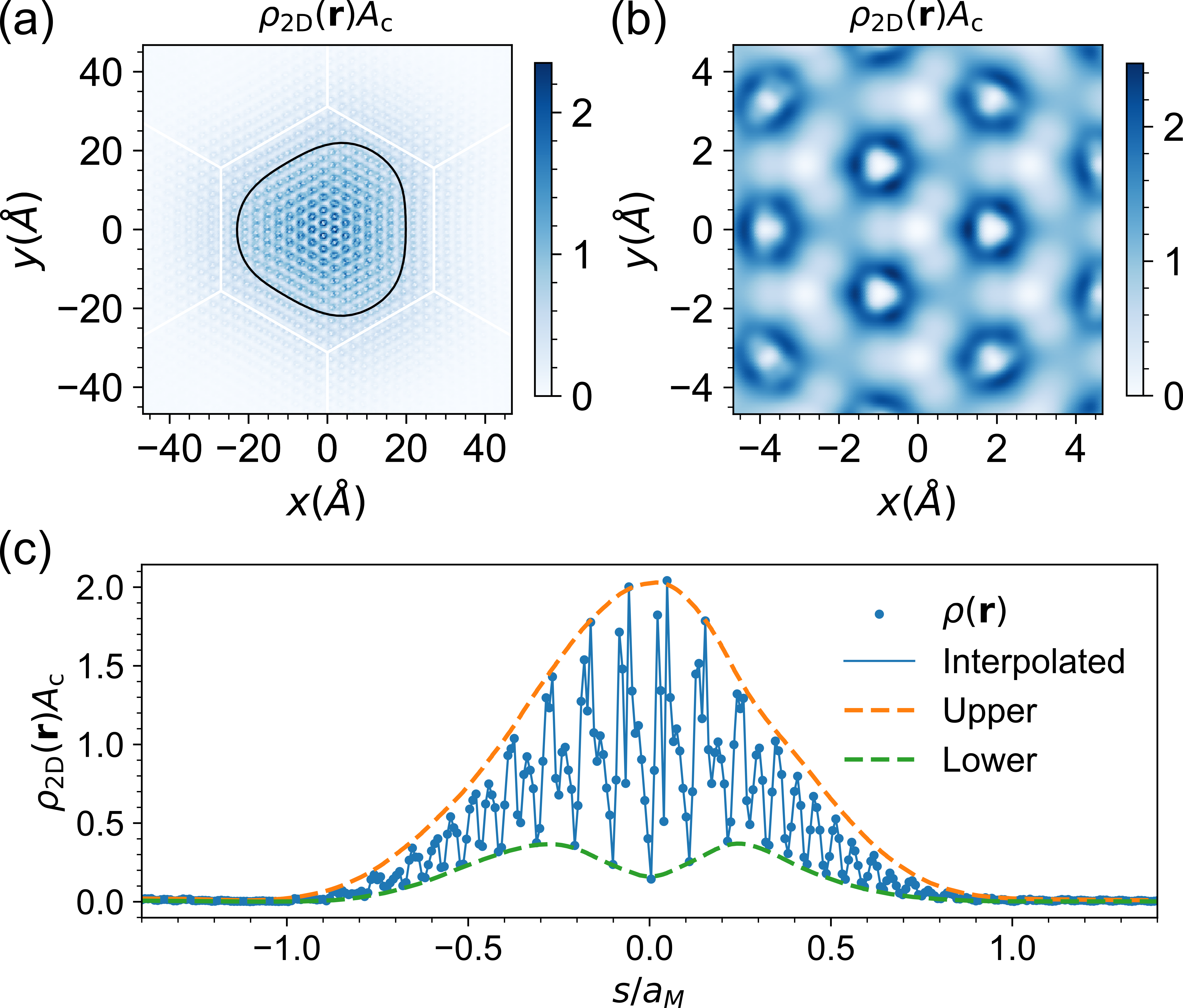}
    \caption{(a) Real space distribution of the $z$-integrated Wannier density $\rho_{\mathrm{2D}}(\mathbf{r})$ at twist angle $\theta=3.5^\circ$. The density is shown in units of $A_c^{-1}$, where $A_c$ is the area of the moiré unit cell. The black contour provides a smoothed guide to the Wannier function's spatial profile, corresponding to half of its maximum value. The white hexagon marks the moiré unit cell.  (b) Zoomed-in view of the same density distribution to atomic scale.  (c) Line cut of $\rho_{\mathrm{2D}}(\mathbf{r})A_c$ along the $\mathbf{a}_1$ direction, plotted as a function of distance $s/a_M$, where $a_M$ is the moiré lattice constant. Blue points show the sampled density, and the solid blue line is the interpolated profile. The dashed orange and green curves denote the upper and lower envelopes, respectively. }
    \label{fig:wannier}
\end{figure}

The resulting Wannier function is shown in Fig.~\ref{fig:wannier}.
The optimized Wannier center is located very close to the $1a$ position of the moiré unit cell, with coordinates $(-6.9\times 10^{-5},-1.15\times 10^{-4})\text{\AA}$, confirming that the orbital is centered at the desired high symmetry position.
On the moiré length scale, the Wannier density is concentrated near the center of one moiré unit cell and exhibits a compact molecular orbital structure.
The localization length of this orbital is on the same order-of-magnitude as the moir\'e lattice constant $a_M \approx 53.996 \text{\AA}$.
Fig.~\ref{fig:wannier}(c) presents a line cut along the $\mathbf{a}_{M1}$ direction in units of $a_M$, together with interpolated upper and lower envelopes.
These envelopes separate the smooth moiré scale profile from the rapid atomic scale oscillations and show that the characteristic spatial extent of the orbital is comparable to the moiré lattice constant.
This spread is consistent with the MLWF constructed from continuum model in its length scale.
While the Wannier function obtained from the continuum model has an approximately isotropic profile, the DFT-derived envelope reflects the reduced lattice symmetry and exhibit a clear $C_3$ rotational symmetry.
A zoomed-in view shown in Fig.~\ref{fig:wannier}(b) further reveals the atomic-scale structure of the Wannier orbital, showing that the smooth moiré scale envelope is modulated by the underlying $\mathrm{WSe_2}$ lattice.
The coexistence of a compact moiré-scale envelope and an atomic-scale internal texture reflects the molecular orbital character of the DFT-derived Wannier function.

This Wannier construction provides a direct bridge between the {\it ab initio} DFT bands and an effective lattice model.
Using the wavefunctions of the MLWF and TPLO, we further compute the hopping and hybridization amplitudes among the $f$ and $c$ orbitals within the DFT-derived model.
Although the $c$ orbital in this model is not exponentially localized, one can still define its dispersion in momentum space and its hybridization with the compact $f$ orbital.
The resulting effective Hamiltonian for the top two moiré bands take the form
\begin{align}
    H_t = \sum_{\mathbf{R}\mathbf{R}'\sigma}t_{ff}^{(\sigma)}&(\mathbf{R}) f^\dagger_{\mathbf{R}+\mathbf{R}', \sigma} f_{\mathbf{R}',\sigma} + \sum_{\vk,\sigma} \varepsilon^{(\sigma)}_c(\vk) c^\dagger_{\vk,\sigma} c_{\vk\, \sigma} \nonumber\\
    + &\sum_{\vk,\sigma}\left( V_{\rm hyb}^{(\sigma)}(\vk) f^\dagger_{\vk,\sigma} c_{\vk,\sigma} + {\rm h.c.}\right)\,.
\end{align}
Here $t^{(\sigma)}_{f}(\mathbf{R})$ describes hopping between the compact $f$ orbitals, $\varepsilon^{(\sigma)}_c(\vk)$ is the effective kinetic energy of the $c$ orbital, and $V_{\rm hyb}^{(\sigma)}(\vk)$ stands for the hybridization between
the $c$ and $f$ orbitals.
For the spin up sector, the nearest neighbor hopping among the $f$ orbitals is given by $|t_{f}| = 4.03 \, \rm meV$, more than 30 times larger than the next-nearest-neighbor hopping, whose relative magnitude is about $|t_{f}'| = 0.12 \, \rm meV$.
Because of the “band inversion” around the $\gamma$ point, the $f$-$c$ hybridization is also sizable; indeed, the maximum value of $|V_{\rm hyb}(\vk)|$ in the MBZ reaches up to $10\ \rm meV$, as shown in Fig.~\ref{fig:hybridization}(b).
The spin down components ($\sigma = \downarrow$) are simply the complex conjugation of their spin up counterparts due to the time-reversal symmetry.
These values are in good agreement with the continuum-model estimates in Fig.~\ref{fig:continuum_tU} and Fig.~\ref{fig:hybridization}(a).

\begin{figure}[t]
    \centering
    \includegraphics[width=\linewidth]{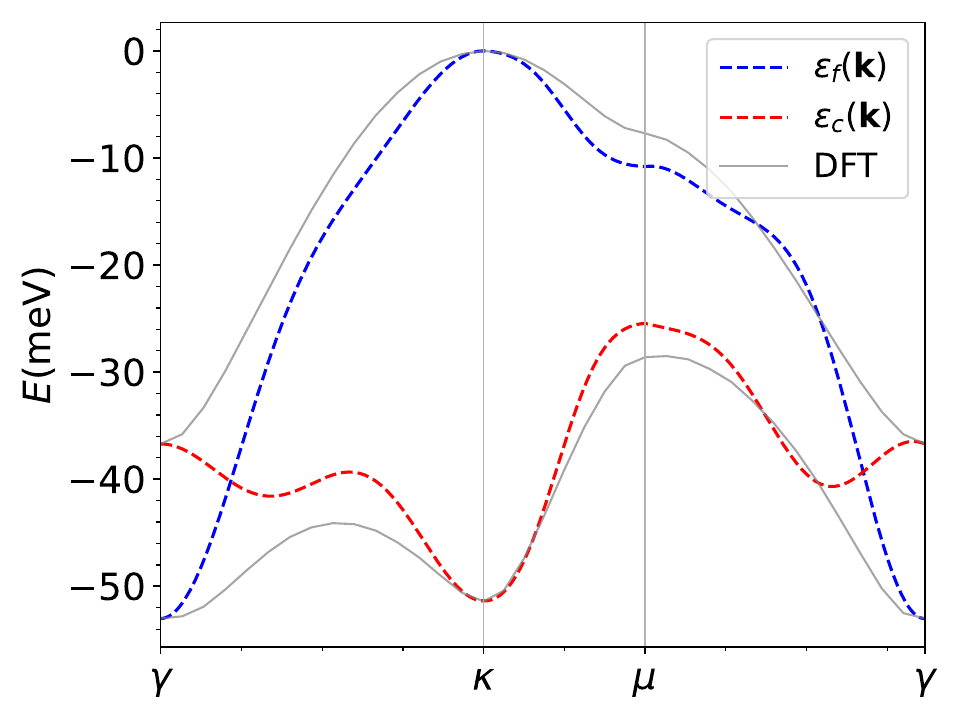}
    \caption{ The band structure of the effective model in the absence of $f$-$c$ hybridization for $\theta=3.5^\circ$. The blue dashed curve denotes the dispersion of the localized $f$ orbital (MLWF), while the red dashed curve stands for the energy of the conduction $c$ band. The grey lines show the corresponding DFT bands. Without hybridization, $f$ and $c$ bands cross near the $\gamma$ point.
    }
    \label{fig:projection}
\end{figure}

In Fig.~\ref{fig:projection}, we present the band structure of the effective model with the hybridization $|V_{\rm hyb}(\vk)|$ switched off.
The blue and red dashed curves represent the dispersion of the localized $f$ orbital described by the MLWF and the itinerant conduction $c$ band, respectively.
In the absence of hybridization, the two bands cross near the $\gamma$ point.
Comparison with the DFT band structure in Fig.~\ref{fig:band}(a) shows that the avoided crossing and band reconstruction near $\gamma$ originate from the hybridization between $f$ and $c$ orbitals.

{\it \color{blue} Coulomb interaction strength}---
We now use the DFT-derived compact molecular orbitals to construct the interaction Hamiltonian.
The dominant direct channel of the interaction elements in the Wannier basis is
\begin{equation}\label{eqn:direct-interaction}
    U(\mathbf{R}) = \frac{1}{2}\int d^2r d^2r'\, U(\mathbf{r} - \mathbf{r}') |W(\mathbf{r} - \mathbf{R})|^2|W(\mathbf{r}')|^2\,,
\end{equation}
where $W(\mathbf{r})$ stands for the 2D-projected Wannier functions of the $f$ orbital (MLWF) and $U(\mathbf{r})$ is the double-gate screened Coulomb interaction.
In the numerical evaluation, we work in momentum space and use the double-gate-screened interaction
\begin{equation}
    U(\mathbf{q}) = \frac{e^2}{2\epsilon_0\epsilon |\mathbf{q}|} \tanh(|\mathbf{q}|d_s)\,.
\end{equation}
Here $d_s = 100\text{\AA}$ is the gate distance and $\epsilon$ is the dielectric constant of the material.
Because the  MLWF is exponentially localized on the moiré scale, the interaction matrix elements are dominated by the on-site Hubbard repulsion, while longer-range and multi-center exchange terms are strongly suppressed.
We therefore retain the leading density-density term of the interaction Hamiltonian in the Wannier basis.
Numerically performing the integral in Eq.~(\ref{eqn:direct-interaction}), we conclude that the on-site interaction $U$ is on the order of magnitude of
$\varepsilon U\sim 640$ meV.
This value is close to the continuum model estimate  $\varepsilon U\sim 500\ \rm{meV}$, as shown in Fig.~\ref{fig:continuum_tU}, providing an additional consistency check between the DFT-derived Wannier construction and the continuum description.

{\it \color{blue} Discussion}---
Our results highlight the importance of topology to the study of correlation physics and, in particular, that the low energy electronic structure of tWSe$_2$ is described by a compact molecular orbital hybridizing with a more extended topological orbital.
The sizable $f$-$c$ hybridization, especially near the moiré $\gamma$ point, quantifies the essential role that the topological component plays for the low-energy effective model.
The DFT-derived parameters can therefore be used to determine the regime of the effective Kondo lattice model \cite{Li2025Topological}.
The $C=(+1, +1)$ topology of the two valence bands is consistent with the structure of chiral $d$ channels proposed in related effective $t$-$J$ model studies, where $d+id$-like pairing naturally emerges from the topological bands.

The real space wave function densities also provide insight into the STM images, which measure the local density of states  \cite{Zhang2020Flat} and could be compared with calculated LDOS patterns constructed from the DFT bands.
The distinct spatial profile of the $f$- and $c$-like components characterizes the orbital character of the low energy states.

{\it \color{blue}Summary---}
We have established the topological nature of
moiré bands pertinent to the superconductivity of tWSe$_2$ in the relevant twist angle range, showing the Chern numbers $C=(+1,+1)$ of the two bands.
Accordingly, we develop a first-principles partial Wannierization framework for the moiré bands,
and use their $C_{3z}$ symmetry representations to
construct a compact molecular orbital together with a more extended topological orbital.
The compact molecular orbitals serve as microscopic building blocks for the effective description of the top valence bands, and provide an {\it ab initio} foundation for
the study of the strongly correlated normal state and unconventional superconductivity in
twisted $\mathrm{WSe_2}$.

\begin{acknowledgments}
{\it Acknowledgments.~}
We thank Lei Chen, Yuan Fang, Kuan-Sen Lin, Kin Fai Mak,  Abhay Pasupathy, Jie Shan, and Shouvik Sur for useful discussions.
This work has been supported in part by
the U.S. Department of Energy, Office of Science, Basic Energy Sciences, under Award No. DE-SC0018197. M.H.N. acknowledges support from the National Science Foundation MRSEC DMR-2308817. J.C. acknowledges the support of the National Science Foundation under Grant No. DMR-1942447, support from the Alfred P. Sloan Foundation through a Sloan Research Fellowship and the support of the Flatiron Institute, a division of the Simons Foundation. R.D. was partially supported by the Texas Quantum Institute Graduate Fellowship. The majority of the computational calculations have been performed on the Shared University Grid at Rice funded by NSF under Grant No.~EIA-0216467, a partnership between Rice University, Sun Microsystems, and Sigma Solutions, Inc., the Big-Data Private-Cloud Research Cyberinfrastructure MRI-award funded by NSF under Grant No. CNS-1338099, and the Advanced Cyberinfrastructure Coordination Ecosystem: Services \& Support (ACCESS) by NSF under Grant No. DMR170109. This work also used computational resources from Stampede3 at The University of Texas at Austin through allocation PHY250206 from the ACCESS program, which is supported by NSF Grants
2138259, 2138286, 2138307, 2137603, and 2138296. F.X., J.C. and Q.S. acknowledge the hospitality of the Aspen Center for Physics, which is supported by NSF grant No. PHY-2210452.
\end{acknowledgments}

\bibliography{reference.bib}
\bibliographystyle{apsrev4-2}

\onecolumngrid
\clearpage

\beginsupplement
\section*{Supplemental Material}
\setcounter{secnumdepth}{3}

\tableofcontents
\section{Computational details of the \textit{ab inito} calculations}
\subsection{Structural reconstructions of the relaxed moir\'{e} superlattices.  }
\begin{figure}[t]
    \centering
    \includegraphics[width=\linewidth]{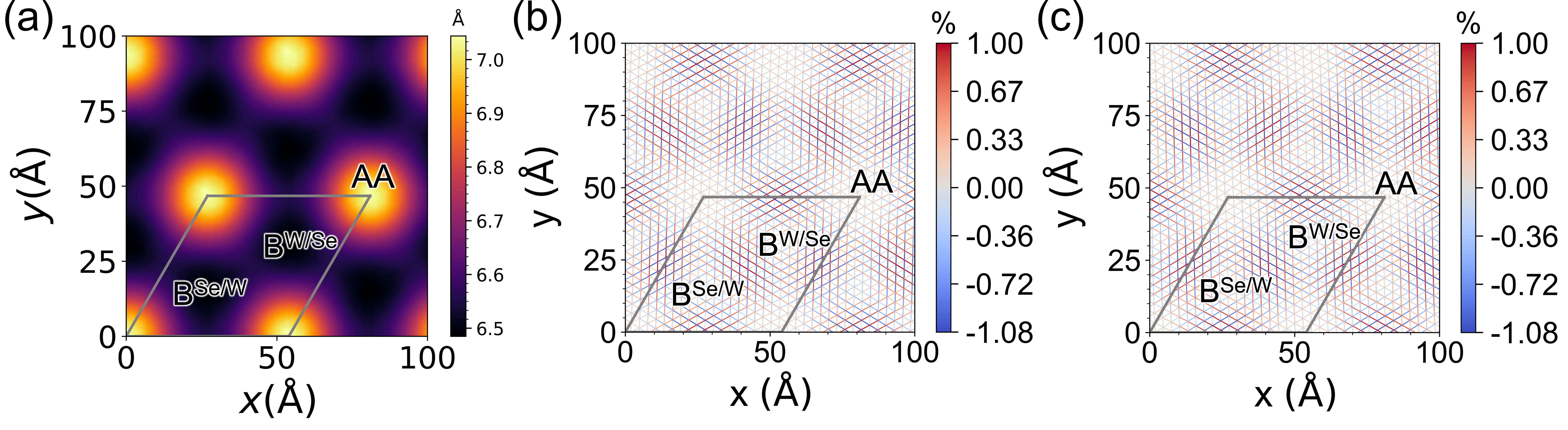}
    \caption{Structural reconstructions of t$\mathrm{WSe_2}$ at twist angle $\theta=3.5^\circ$ (a) Distribution of the interlayer spacing in t$\mathrm{WSe_2}$ for $\theta=3.5^\circ$ measured as the local interlayer distance between W atoms. (b) Strains for the top layer.  (c) Strains for the bottom layer. The local in-plane strain is defined on each W-W nearest neighbor segment as the relative change from the equilibrium lattice constant of 3.28 \AA.}
    \label{fig:reconstruction}
\end{figure}

The superlattices were generated using the TWISTER code~\cite{naik2022twister} and relaxed using classical force fields parameterized against van der Waals corrected DFT data. Intralayer and interlayer interactions were described with Stillinger--Weber~\cite{stillinger1985computer} and Kolmogorov--Crespi~\cite{kolmogorov2005registry, Naik2019Kolmogorov}  potentials, respectively, as implemented within the LAMMPS package~\cite{thompson2022lammps}.
We used the experimental lattice constant of 3.28~\AA{} for WSe$_2$. The structural reconstruction of the moiré superlattice is shown in Fig.~\ref{fig:reconstruction}.

\subsection{Density functional theory calculations}
The single-particle electronic structure of the moiré superlattices was investigated within density functional theory (DFT)~\cite{kohn1965self} using the SIESTA code~\cite{soler2002siesta}. The calculations employed fully relativistic norm-conserving pseudopotentials~\cite{van2018pseudodojo, hamann2013optimized} with spin-orbit coupling, and the Perdew-Burke-Ernzerhof generalized-gradient functional~\cite{perdew1996generalized} for exchange and correlation. The ground-state charge density was first converged self-consistently by sampling the moiré $\bm{\gamma}$-point. This density was then used to construct and diagonalize the DFT Hamiltonian on a denser set of $k$-points across the moiré Brillouin zone (BZ), yielding the band structure and wavefunctions used as input to subsequent calculations.

\subsection{Chern numbers}
The Kohn-Sham wavefunctions from SIESTA are converted from the numerical
atomic-orbital basis to a plane-wave basis with a $40$~Ry energy cutoff. The
four topmost moir\'e valence bands form two twofold-degenerate, time-reversed
groups $v_1$ and $v_2$. Because spin-orbit coupling mixes $\uparrow$ and
$\downarrow$, we diagonalize the spin operator $\sigma_z$ projected onto each
degenerate subspace, yielding maximally spin-polarized bands. At the three
$C_3$-invariant momenta of the moir\'e Brillouin zone, we extract
the threefold-rotation eigenvalues $\theta_\gamma$, $\theta_\kappa$, and
$\theta_{\kappa'}$. The Chern number $C$ of each isolated spin-diagonalized band
follows from
$\exp(i\tfrac{2\pi}{3}C)=-\,\theta_\gamma\theta_\kappa\theta_{\kappa'}$, which
fixes $C$ modulo $3$~\cite{PhysRevB.86.115112}.

\section{Wannierization from DFT bands}
\label{sec:Wannierization_DFT}
In this section, we will briefly describe the procedure used to construct compact molecular orbitals from the DFT bands.
We focus on the spin-up sector originating from the $K$-valley and start from an initial trial function whose charge center and symmetry quantum numbers match those of the desired compact molecular orbital.
The trial orbital is centered at the $1a$ Wyckoff position and is chosen to transform properly under the $C_{3z}$ symmetry
\begin{align}
    \Phi_{\mathrm{trial}}=\frac{1}{\sqrt{N_{\mathbf{K}_0}}}\sum_{\mathbf{K}_0} \Phi_l(\vk-\mathbf{K}_0)\,,
\end{align}
where $N_{\mathbf{K}_0}$ is the number of $\mathbf{K}_0$ included in the sum and $\Phi_l(\vk)$ is the Fourier transformation of the Gaussian wave packet with integer angular momentum $l\geq0$ in two dimensional space
\begin{align}
    \Phi_l(\vk)=2\sqrt{\frac{\pi}{|l|!}}\left(\frac{1}{\sigma_\vk}\right)^{|l|+1}e^{-\frac{\vk^2}{2\sigma_\vk^2}}(-1)^l(k_x+ik_y)^l\,.
\end{align}
We choose $\hat{\mathbf{K}}_0=(0,-1),~(\frac{\sqrt{3}}{2},\frac{1}{2}),~(-\frac{\sqrt{3}}{2},\frac{1}{2})$ with $|\mathbf{K}_0|=1.2\mathrm{\AA}^{-1}$ and $\sigma_\vk=0.6\mathrm{\AA}^{-1}$.
This choice preserves $C_{3z}$ symmetry and enhances the overlap between the trial orbital and the DFT Bloch states.
For $l=1$, the trial orbital carries the desired   $C_{3z}$ eigenvalue, satisfying
\begin{align}
    \Phi_{l=1}(C_{3z}\vk)=e^{i\frac{2\pi}{3}}\Phi_{l=1}(\vk)\,.
\end{align}
We evaluate this trial orbital on the $6\times 6$ momentum grid and provide the resulting projection data as input to \textsc{wannier90}.

The \textsc{wannier90} calculation determines a smooth $\vk$-dependent linear combination of the top two moiré bands that minimizes the spread of the resulting compact orbital.
In terms of the creation operator $c_{\vk,n}^\dagger$ for the two bands, the Bloch wave function of the MLWF is written as
\begin{align}
    f_\vk^\dagger=\sum_{n=1,2}c_{\vk,n}^\dagger U_{n,f}^{\mathrm{dis}}(\vk)\,,
\end{align}
where $U_{n,f}^{\mathrm{dis}}(\vk)$ is the output rectangular "disentanglement matrix".
The remaining orbital, denoted as the conduction orbital $c$, is then constructed by a Gram–Schmidt process with a total Chern number $C=+2$ \cite{Xie2025Kondo}.

\begin{figure}[t]
    \centering
    \includegraphics[width=0.7\linewidth]{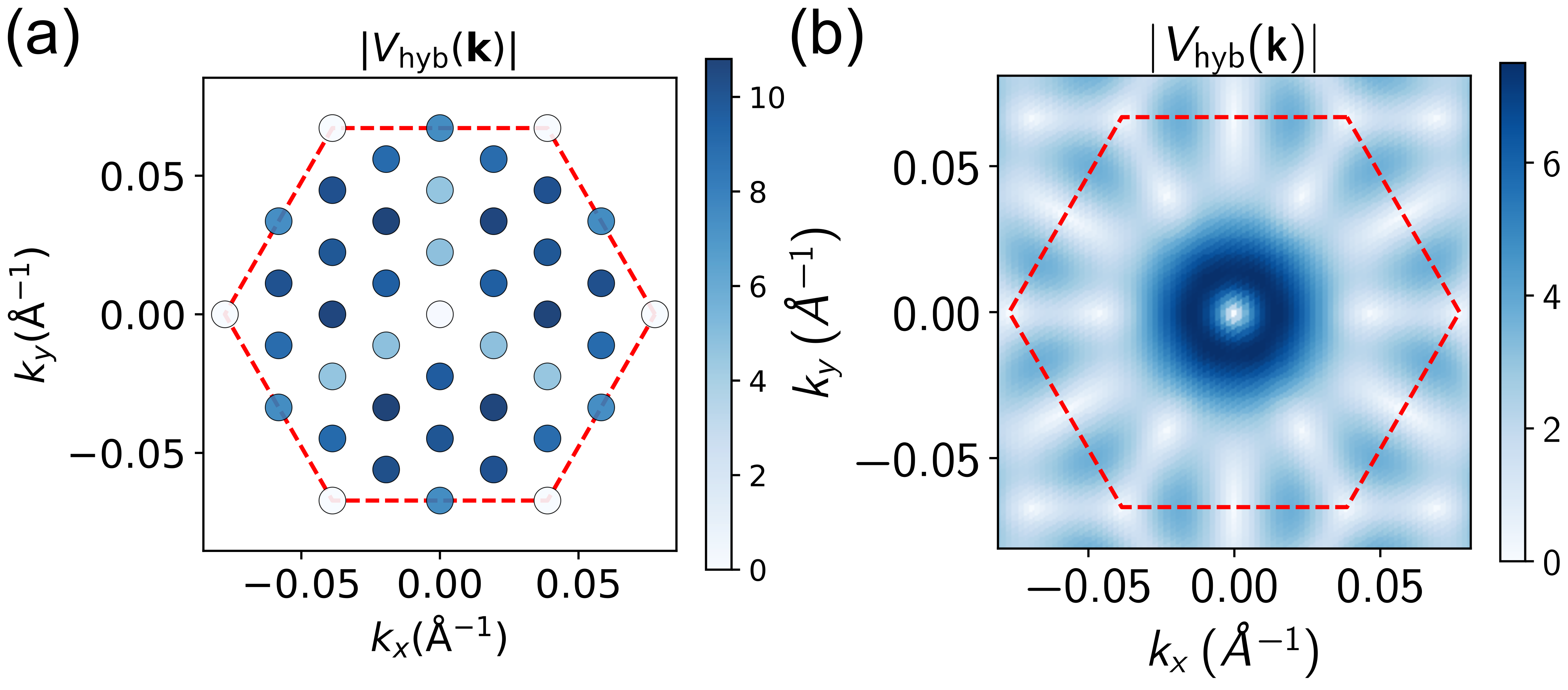}
    \caption{Absolute value of the hybridization function, $|V_{\mathrm{hyb}}(\mathbf{k})|$, over the moiré Brillouin zone (MBZ).  (a) Result from the DFT calculations at $3.5^\circ$.  (b) Result from the continuum model, using the parameters specified in Sec.~\ref{sec:Wannierization_cont}. The red dashed hexagon stands for the boundary of the MBZ.}
    \label{fig:hybridization}
\end{figure}

After obtaining the $f$ and $c$ wave functions,
We can then project the Hamiltonian into the $2\times 2$ Hilbert space spanned by the two orbitals and evaluate the matrix elements numerically.
This gives the effective $f$ orbital hopping, $c$ band dispersion and the hybridization $V_\mathrm{hyb}(\vk)$ between the two orbitals.
Fig.~\ref{fig:hybridization}(a) shows the absolute value of the hybridization $V_\mathrm{hyb}(\vk)$ in the first Brillouin zone.
The hybridization is very small at the moiré $\gamma$ and $\kappa$ points, but reaches a maximal value of approximately $10$ meV away from these high-symmetry points.
This momentum dependence reflects the band inversion and orbital mixing in the system.

\section{Continuum model and Wannierization}
\label{sec:Wannierization_cont}
In this section, we briefly outline the properties of the continuum model, as well as the Wannierization procedure that we adopted in Ref.~\cite{Xie2025Kondo}.
It is widely considered that the low energy electronic states in single layer WSe$_2$ is dominated by the valence band edges around $K$ and $K'$ points, that can be well-captured by a quadratic dispersion.
When the two layers of WSe$_2$ are stacked on top of each other, the electrons will experience intra-layer potentials and inter-layer tunnelings, which are periodic on the moir\'e lattice scale.
These electronic states can be described by the following continuum model \cite{Wu2019topological}:
\begin{equation}
    h(\mathbf{r}, -i\bm{\nabla}) = \left(\begin{matrix}
            \frac{\bm{\nabla}^2}{2m^*} + \tilde{v}_+(\mathbf{r}) + \frac{\varepsilon_D}{2}  & T(\mathbf{r}) \\
            T^*(\mathbf{r}) & \frac{\bm{\nabla}^2}{2m^*} + \tilde{v}_-(\mathbf{r}) - \frac{\varepsilon_D}{2}
\end{matrix}\right)\,.
\end{equation}
Here, the two entries of this Hamiltonian stands for the top and bottom layers, and $m^*$ is the effective mass of the single-layer band edge states around the $K$ and $K'$ points.
The intra- and inter-layer potentials have the form of:
\begin{align}
    \tilde{v}_\pm(\vlr) =& 2\tilde{v} \sum_{j = 1}^3 \cos(  \mathbf{g}_j \cdot \vlr \pm \psi )\,,\\
    T(\vlr) =& w \sum_{j=1}^3 e^{i\vq_j\cdot \vlr}\,,
\end{align}
where $\mathbf{g}_1 = \mathbf{b}_1$, $\mathbf{g}_2 = \mathbf{b}_2$, $\mathbf{g}_3 = -\mathbf{b}_1-\mathbf{b}_2$, and $\vq_1 = \frac13 \mathbf{b}_1 + \frac23 \mathbf{b}_2$, $\vq_2 = \vq_1 - \mathbf{b}_1 - \mathbf{b}_2$, $\vq_3 = \vq_1 -\mathbf{b}_2$.
There are various parameter choices that has been proposed for the continuum model in the literature \cite{Wu2019topological, Pan2020Band, Devakul2021Magic, Mao2024Transfer, Xu2025Signatures, Xu2025Chiral}.
Here we use the results from Ref.~\cite{Devakul2021Magic}, in which the parameters are chosen as $(m^*, \tilde{v}, \psi, w) = (0.43 m_e, 9{\rm \, meV}, 128^\circ, 18{\rm \, meV})$.

\begin{figure}[t]
    \centering
    \includegraphics[width=0.8\linewidth]{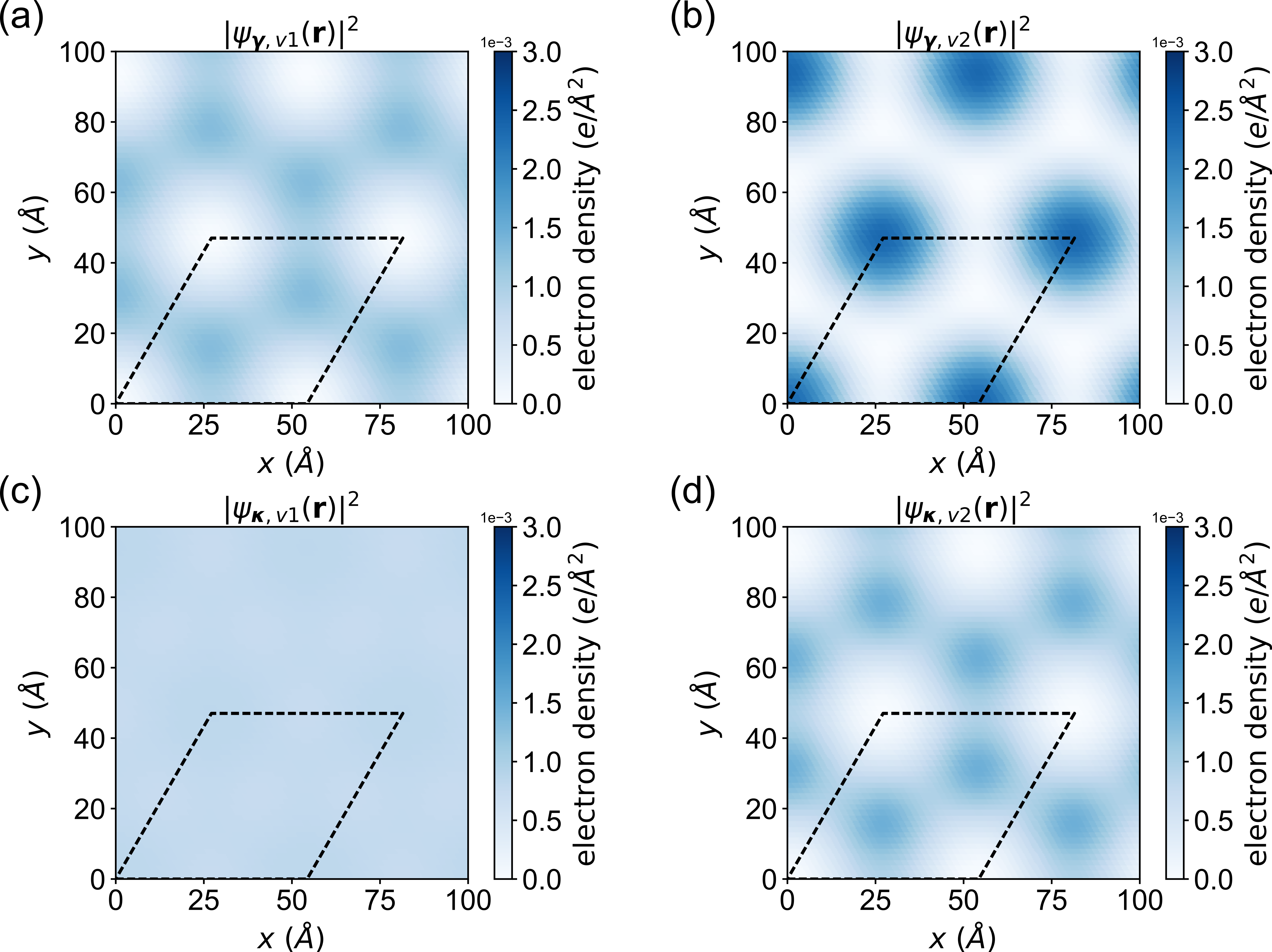}
    \caption{Real space electron density (both spins) at twist angle $\theta=3.5^\circ$. Panels (a,b) are evaluated at the moir\'e $\gamma$ point, and panels (c,d) at the moir\'e $\kappa$ point. Panels (a,c) show the topmost band, while panels (b,d) show the second top band. Dashed parallelgrams mark one moir\'e unit cell. }
    \label{fig:cont_density}
\end{figure}

\begin{figure}[t]
    \centering
    \includegraphics[width=0.8\linewidth]{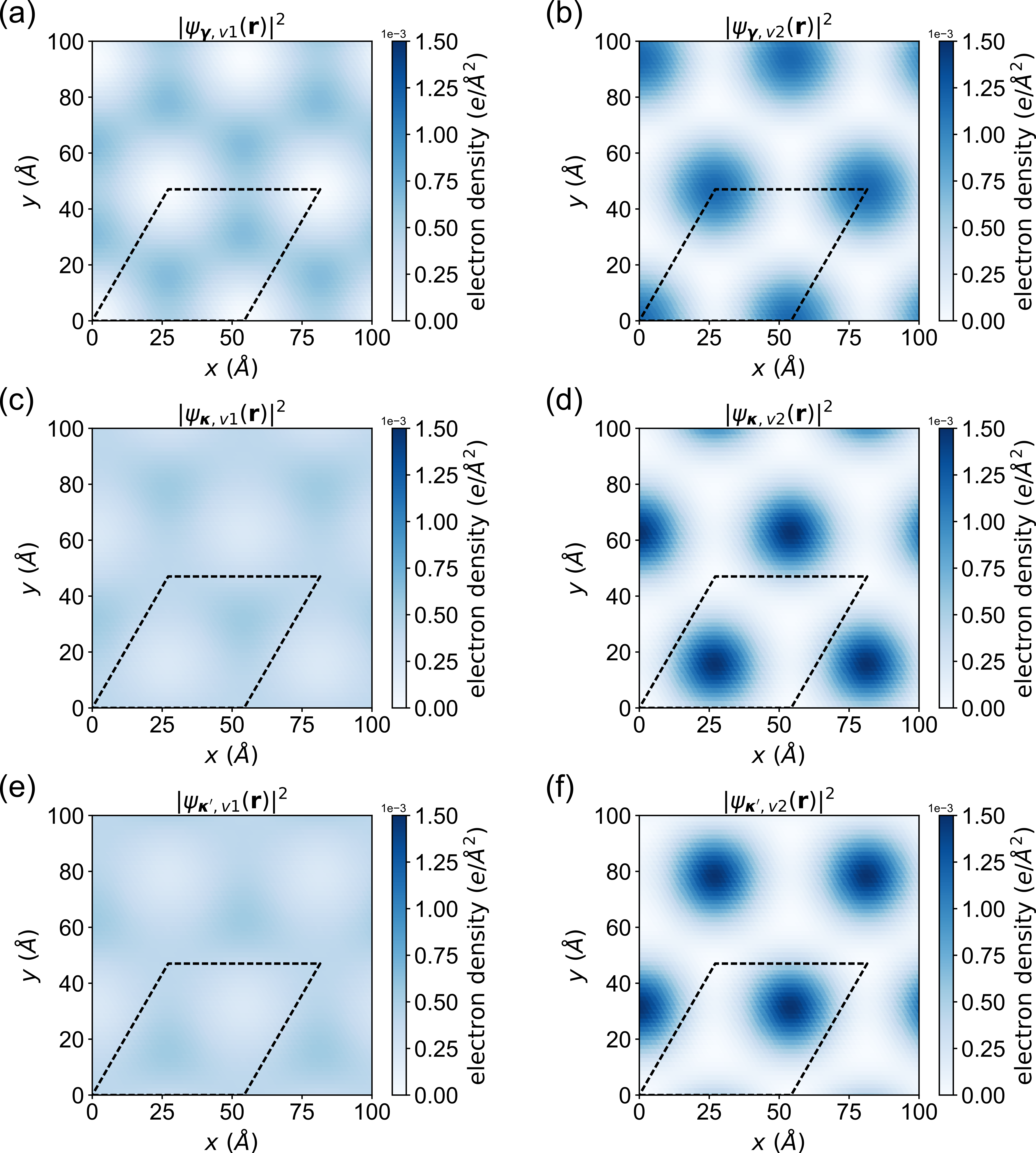}
    \caption{Real space electron density (single spin) at twist angle $\theta=3.5^\circ$. Panels (a,b) are evaluated at the moir\'e $\gamma$ point, panels (c,d) at the moir\'e $\kappa$ point, and panels (e,f) at the moir\'e $\kappa'$ point. Panels (a,c) show the topmost band, while panels (b,d) show the second top band. Dashed parallelgrams mark one moir\'e unit cell. }
    \label{fig:cont_density_singlespin}
\end{figure}

The eigenstates of the continuum model can be solved by representing the Hamiltonian in the plane wave basis with the following form:
\begin{equation}
    |\psi_{\vk n}\rangle = \sum_{\mathbf{Q}} u_{\mathbf{Q},n}(\vk) |\vk - \mathbf{Q}\rangle\,,
\end{equation}
where $\mathbf{Q}$ stands for the reciprocal wave vectors of both top and bottom layers.
Using the aforementioned model parameters at twisting angle $\theta = 3.5^\circ$, the top two moir\'e bands carry the same valley Chern number $C = (+1, +1)$, as we have seen in the {\it ab initio} calculations.
The density distribution of the these Bloch states at $\gamma$ and $\kappa$ points in the moir\'e Brillouin zone can be found in Figs.~\ref{fig:cont_density} and \ref{fig:cont_density_singlespin}.

\begin{figure}[t]
    \centering
    \includegraphics[width=\linewidth]{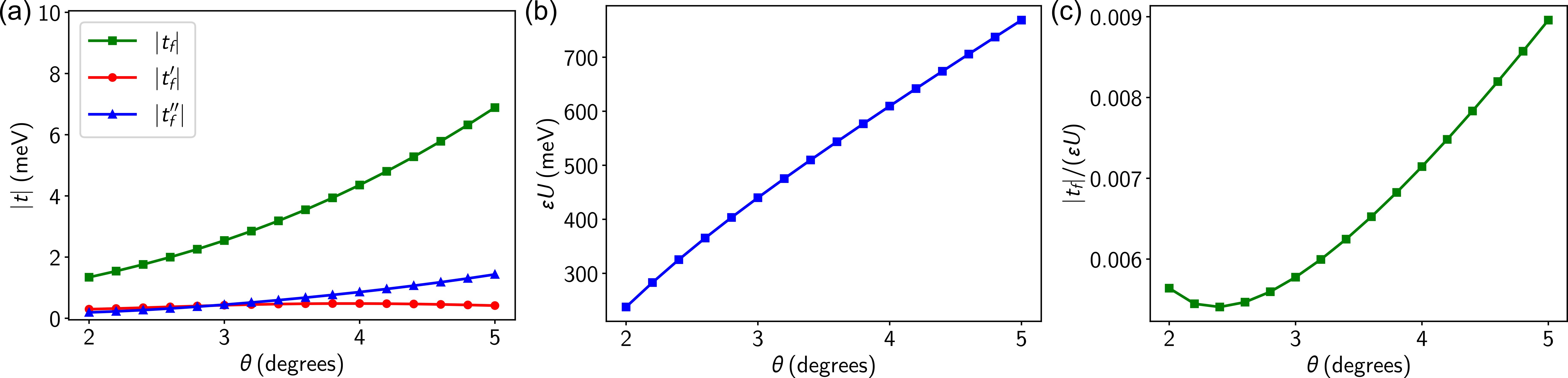}
    \caption{(a) Hoppings of the $f$ orbital as a function of twist angle $\theta$. (b) On-site $U$ as a function of $\theta$. (c) The ratio of $t_f/U$ increases with twist angle $\theta$.
    }
    \label{fig:continuum_tU}
\end{figure}

Performing a similar partial Wannierization procedure as outlined in the previous section, we are able to construct the maximally localized Wannier function $f$, which also hybridize with a topological power-law orbital $c$ that carries a non-vanishing valley Chern number $C = 2$.
Using the constructed $f$ orbital from the continuum model, we are also able to evaluate the nearest, next-nearest and next-next-nearest hoppings $t, t', t''$, as well as the on-site repulsive Coulomb interactions $\varepsilon U$ as a function of twisting angle.
The results can be found in Fig.~\ref{fig:continuum_tU}.
Besides, the hybridization function between the two orbitals, shown in Fig.~\ref{fig:hybridization} is also computed from this Wannierization procedure.

\end{document}